\documentclass[aps,prc,twocolumn,superscriptaddress,showpacs,floatfix,nofootinbib]{revtex4-1}
\usepackage{amsmath,graphicx,color}
\usepackage[vcentermath]{youngtab}
\begin{document}

\title{Systematic procedure for analyzing cumulants at any order}

\author{Philippe Di Francesco}
\affiliation{Institut de physique th\'eorique, Universit\'e Paris Saclay, CNRS, CEA, F-91191 Gif-sur-Yvette, France} 
\affiliation{Department of Mathematics, University of Illinois MC-382, Urbana, IL 61821, U.S.A. }
\author{Maxime Guilbaud}
\affiliation{Rice University, Houston, USA}
\author{Matthew Luzum}
\affiliation{Instituto de F\'\i sica, Universidade de S\~ao Paulo, C.P. 66318, 05315-970 S\~ao Paulo, SP, Brazil}  
\author{Jean-Yves Ollitrault}
\affiliation{Institut de physique th\'eorique, Universit\'e Paris Saclay, CNRS, CEA, F-91191 Gif-sur-Yvette, France} 

\date{\today}

\begin{abstract}
We present a systematic procedure for analyzing cumulants to arbitrary order in the context of heavy-ion collisions. 
It generalizes and improves existing procedures in many respects. In particular, particles which are correlated are allowed to belong to different phase-space windows, which may overlap.  It also allows for the analysis of cumulants at any order, using a simple algorithm rather than complicated expressions to be derived and coded by hand. In the case of azimuthal correlations, it automatically corrects to leading order for detector non-uniformity, and it is useful for numerous other applications as well.
We discuss several of these applications: anisotropic flow, event-plane correlations, symmetric cumulants, net baryon and net charge fluctuations. 
\end{abstract}
\maketitle

\section{Introduction}

A nucleus-nucleus collision at ultrarelativistic energies typically emits thousands of particles~\cite{Abbas:2013bpa}. 
The large multiplicity enables one to accurately measure various high-order correlations and cumulants. 
Cumulants are {\it connected\/} correlations, and high-order cumulants are typically used to probe the emergence of collective effects. 
More specifically, cumulants of azimuthal correlations have been 
measured~\cite{Adler:2002pu,Alt:2003ab,Aamodt:2010pa,Chatrchyan:2012ta,Aad:2014vba}
in order to study the collective expansion of the quark-gluon plasma in the direction of the impact parameter, or elliptic flow~\cite{Ollitrault:1992bk,Heinz:2013th}. 
Cumulants of the net proton~\cite{Aggarwal:2010wy} and net charge~\cite{Adamczyk:2014fia,Adare:2015aqk} distribution 
are analyzed in order to search for the critical fluctuations associated with the QCD phase 
transition at finite baryon density~\cite{Stephanov:2008qz,Gupta:2011wh}. 
Recent years have witnessed the emergence of a great variety of new correlation observables: 
Event-plane correlations~\cite{ALICE:2011ab,Aad:2014fla},
correlations between transverse momentum  
and anisotropic flow~\cite{Bozek:2016yoj}, or between two different Fourier harmonics of anisotropic flow~\cite{Bilandzic:2013kga,ALICE:2016kpq}, which all involve cumulants of multiparticle correlations. 

We propose a new framework for the cumulant analysis, which is more flexible than existing frameworks and allows to fully exploit the potential of multiparticle correlation analyses in high-energy collisions. 
Our understanding of the collision dynamics has considerably evolved since cumulants were introduced in this context~\cite{Borghini:2000sa}. 
The importance of event-to-event fluctuations in the flow pattern~\cite{Aguiar:2001ac,Miller:2003kd} was only recognized later~\cite{Alver:2006wh,Alver:2010gr}. 
More specifically, it was recently shown that the rapidity dependence of anisotropic flow fluctuates event to event~\cite{Bozek:2010vz,Khachatryan:2015oea}.
This longitudinal decorrelation~\cite{Pang:2014pxa} induces a sizable variation of azimuthal 
correlations with the relative rapidity~\cite{Adamczyk:2013waa}. 
Existing analysis frameworks~\cite{Borghini:2001zr,Bilandzic:2010jr},
where all particles but one\footnote{The analysis of differential flow
  with cumulants correlates one particle from a restricted phase space
  window with reference particles which are all in the same window.}
are taken from the same rapidity window, do not allow to study this
effect.  

Precision studies of high-order cumulants are also needed. They have been argued to be a crucial probe of collective behavior in proton-nucleus collisions~\cite{Aad:2013fja,Chatrchyan:2013nka,Yan:2013laa,Abelev:2014mda} and proton-proton collisions~\cite{Khachatryan:2016txc,Aad:2015gqa} and first analyses of order 6 and 8 cumulants are promising~\cite{Khachatryan:2015waa}. 
High-order cumulants also provide insight into non-Gaussian fluctuations in nucleus-nucleus collisions~\cite{Bhalerao:2011yg,Gronqvist:2016hym,Giacalone:2016eyu}, where they have also been measured up to order 8~\cite{Aad:2014vba,Abelev:2014mda}. 
The maximum value of 8 is dictated by existing analysis frameworks, but higher orders are feasible experimentally.  

The oldest framework~\cite{Borghini:2001vi} extracts cumulants by numerically tabulating the generating function and using a finite-difference approximation to compute its successive derivatives. The numerical errors resulting from this procedure are hard to evaluate. 
A new framework by Bilandzic \textit{et al.}~\cite{Bilandzic:2010jr}
uses explicit expressions of cumulants in terms of moments of the
flow vector, which is a more robust approach. However, only a finite set of cumulants are provided, and azimuthal symmetry is assumed in order to simplify the algebraic complexity of these expressions. Azimuthal asymmetries in the detector acceptance and efficiency must therefore be corrected beforehand. On the other hand, the cumulant analysis automatically corrects for such asymmetries~\cite{Borghini:2000sa}, so it is tempting to use cumulants for this purpose as well. 

Our new framework generalizes the approach of Bilandzic~\cite{Bilandzic:2010jr} in several respects:
\begin{itemize}
\item
It applies to arbitrary observables, not only to Fourier coefficients of the azimuthal distribution.
\item
One can correlate particles from different bins in rapidity and transverse momentum. 
\item
No assumption is made regarding the detector acceptance and efficiency.
\item 
Cumulants can be evaluated to arbitrarily high order. 
\end{itemize}
In Sec.~\ref{s:pairs}, we illustrate the procedure on the simple example of azimuthal pair correlations. 
We then generalize results to higher-order correlations in Sec.~\ref{s:selfcorrelations}, where we show how to remove self-correlations to all orders~\cite{Bilandzic:2013kga}, and in 
Sec.~\ref{s:cumulants}, where we give the inversion formulas for cumulants as a function of moments to all orders~\cite{kubo}.
Specific implementations are discussed in Sec.~\ref{s:applications}. 

\section{Example: pair correlation}
\label{s:pairs}

We illustrate the steps of the calculation on the simple example of azimuthal pair correlations~\cite{Wang:1991qh}.
One typically wants to compute an average value of $\cos n\Delta\Phi$, where $n$ is the harmonic order and $\Delta\Phi$ is the relative azimuthal angle between two particles in the same event~\cite{Aamodt:2011by}.
Specifically, one takes the first particle from a region of momentum space $A$ and the second from region $B$, and one evaluates
\begin{equation}
\label{defpaircorr}
\langle\cos n\Delta\Phi\rangle\equiv \frac{\left\langle\sum_{\rm
    pairs} e^{in(\phi_j-\phi_k)}\right\rangle}{\left\langle\sum_{\rm pairs} \phantom{mm}1\phantom{mm}\right\rangle},
\end{equation}
where $\phi_j$ and $\phi_k$ are the azimuthal angles of particles
belonging to the same event, 
``pairs'' is a shorthand notation for ``$j\in A$, $k\in
B$ and $j\not= k$'', 
and angular brackets in the
right-hand side denote an average over events.\footnote{The imaginary
  part of the right-hand side vanishes if parity is conserved, up to
  statistical fluctuations and asymmetries in the detector efficiency.} 
  
One could evaluate the sums over pairs as nested loops over $j$ and $k$, but it is more efficient to instead factorize the sums~\cite{Bilandzic:2010jr}.  For example, if $A$ and $B$ are disjoint:
\begin{equation}
\label{factorization}
\sum_{\rm pairs} e^{in(\phi_j-\phi_k)}=\sum_{j\in A}e^{in\phi_j} \sum_{k\in B}e^{-in\phi_k}. 
\end{equation}
In the case where regions $A$ and $B$ overlap, such that they share some of the same particles, one must exclude the extra terms with $j=k$, corresponding to a trivial correlation of a particle with itself (self-correlation): 
\begin{equation}
\label{selfcorr2}
\sum_{\rm pairs} e^{in(\phi_j-\phi_k)}=\sum_{j\in A}e^{in\phi_j} \sum_{k\in B}e^{-in\phi_k}-\sum_{j\in A\cap B} 1,
\end{equation}
where the final sum is over the intersection
of sets $A$ and $B$, $A\cap B$, and each term has unit contribution
since $e^{in(\phi_j - \phi_j)} = 1$  

The next step is to average over a large number of collision events. 
We first assume that particles are independent, in the sense that the
number of particles in two disjoint momentum bins ${\bf p}_1$ and
${\bf p}_2$ are independent variables. Then, the average number of
pairs factorizes as a product of single averages, and 
\begin{equation} 
\label{acceptance2}
\left\langle\sum_{\rm pairs} e^{in(\phi_j-\phi_k)}\right\rangle =
\left\langle\sum_{j\in A} e^{in\phi_j}\right\rangle
\left\langle\sum_{k\in B} e^{-in\phi_k}\right\rangle.
\end{equation}
For an ideal (isotropic) detector and azimuthal symmetric regions $A$
and $B$, the right-hand side vanishes identically, since every
collision event has an arbitrary azimuthal orientation.
In a more realistic experimental situation, the detector efficiency has azimuthal asymmetries and the right-hand side is non-zero. 
However, Eq.~(\ref{acceptance2}) is still valid when particles are independent.

In the more general case where particles are not independent, 
we define the connected correlation as the difference between the two
sides of this equation. It 
thus isolates the physical correlation, and 
naturally corrects for
asymmetries in the detector:
\begin{eqnarray} 
\label{cumulant2}
\left\langle\sum_{\rm pairs}
e^{in(\phi_j-\phi_k)}\right\rangle_c&\equiv& 
\left\langle\sum_{\rm pairs} e^{in(\phi_j-\phi_k)}\right\rangle\cr
&&-\left\langle\sum_{j\in A} e^{in\phi_j}\right\rangle
\left\langle\sum_{k\in B} e^{-in\phi_k}\right\rangle, \cr
&&
\end{eqnarray}
where the subscript $c$ in the left-hand side denotes the connected part of the correlation, i.e., the cumulant~\cite{Borghini:2000sa}.
Note, however, that a non-uniform efficiency introduces a
``cross-harmonic bias''~\cite{Hansen:2014pwa} and the cumulant
involves in general several harmonics of the particle distribution,
not just harmonic $n$~\cite{Bhalerao:2003xf}. 

Note that our definition of independent particles is not exactly
equivalent to assuming that for a given pair of particles, azimuthal
angles $\phi_j$ and $\phi_k$ are independent~\cite{Bilandzic:2010jr}. 
With this alternative definition, one can write an equation similar to
Eq.~(\ref{acceptance2}), where the left-hand side is divided by the
average number of pairs, and the single-particle averages in the
right-hand side are divided by the average number of particles in $A$ and
$B$. 
The definition of the cumulant is then modified accordingly:
\begin{eqnarray} 
\label{cumulantante}
\langle\cos n\Delta\Phi\rangle
&\equiv& 
\frac{\left\langle\sum_{\rm pairs} e^{in(\phi_j-\phi_k)}\right\rangle}
{\left\langle\sum_{\rm pairs}\phantom{mm} 1\phantom{mm}\right\rangle}
\cr
&&-
\frac{\left\langle\sum_{j\in A} e^{in\phi_j}\right\rangle}{\left\langle\sum_{j\in A}\phantom{m} 1\phantom{m}\right\rangle}
\frac{\left\langle\sum_{k\in B} e^{-in\phi_k}\right\rangle}{\left\langle\sum_{k\in B} \phantom{m}1\phantom{m}\right\rangle}. \cr
&&
\end{eqnarray}
The method in this paper can be applied with either choice, 
as explained at the end of Sec.~\ref{s:cumulants}. 
The advantage of our definition is that it simpler, and 
allows to treat multiplicity fluctuations and correlations on the same
footing as anisotropic flow. 
Its apparent drawback is that a correlation can be induced by a 
large fluctuation of the global multiplicity, which is of 
little physical interest. 
However, one easily gets rid of this effect by using a fine centrality  
binning, which is recommended anyway for correlation analyses~\cite{Gardim:2016nrr}.\footnote{Note that early cumulant analyses, which used wide centrality bins due to limited statistics, used a fixed subset of the multiplicity~\cite{Alt:2003ab} in order to avoid the effects of multiplicity fluctuations in conjunction with detector asymmetries and nonflow correlations.}

Instead of explicitly considering the connected correlation
Eq.~(\ref{cumulant2}) (or its variant Eq.~(\ref{cumulantante})) 
to correct for detector anisotropy, one can do the correction in other ways.
Bilandzic \textit{et al.}~\cite{Bilandzic:2010jr} 
weight each particle with $1/p$, where $p$ is 
the efficiency (probability of detection) at the point where the
particle hits the detector. 
After the correction, azimuthal asymmetry can be assumed, and the
first term alone in the right-hand side of Eq.~(\ref{cumulant2}) is equivalent
to the connected part.
This ``inverse weighting'' method can 
still be used here, but is no longer necessary. In particular, note that
inverse weighting cannot be applied when there are holes in the
detector, in which case the efficiency $p$ is $0$. 

Eliminating the nested sums gives the final expression for the numerator and denominator of the measurement
\begin{align}
\left\langle\sum_{\rm pairs} e^{in(\phi_j-\phi_k)}\right\rangle_c\equiv&
\left\langle\sum_{j\in A}e^{in\phi_j} \sum_{k\in B}e^{-in\phi_k}\right\rangle \nonumber \\
&-\left\langle\sum_{j\in A\cap B} 1\right\rangle \nonumber \\
&-\left\langle\sum_{j\in A} e^{in\phi_j}\right\rangle
\left\langle\sum_{k\in B} e^{-in\phi_k}\right\rangle.
\end{align}
and
\begin{align}
\left\langle\sum_{\rm pairs} 1\right\rangle &= 
 \left\langle\sum_{j\in A} 1 \sum_{k\in B} 1\right\rangle -\left\langle\sum_{j\in A\cap B} 1\right\rangle .
\end{align}

In the following Sections, we generalize the above discussion to higher-order correlations and arbitrary observables.

\section{Subtracting self-correlations}
\label{s:selfcorrelations}

Cumulants can be constructed from moments, which are correlations that count $n$-tuples in a collision event, where $n$ now denotes the order of the correlation. 
Depending on the observable, one may weight every particle differently depending on its momentum, as in Eq.~(\ref{defpaircorr}). 
In general, one evaluates in each event multiple sums of the type  
\begin{equation}
\label{sumalldifferent}
Q(A_1,\ldots,A_n)\equiv\sum_{j_1\in A_1,\ldots,j_n\in
  A_n}q_1(j_1)\ldots q_n(j_n),
\end{equation} 
where $j_1,\cdots,j_n$ label particles chosen from $n$ sets labelled $A_1,\cdots,A_n$ (representing, e.g., specific regions in $p_T$ and $\eta$, or particular particle species), all indices in the sum are different, 
and $q_i(j)$ are functions of the particle momentum.  
In the numerator of Eq.~(\ref{defpaircorr}), for example, we have
 $q_1(j) = e^{in\phi_j}$ and $q_2(j) = e^{-in\phi_j}$.

The sum runs over all possible $n$-tuples in the same event. 
If one uses nested loops, the time needed to evaluate such sums grows with the multiplicity $M$ like $M^n$, which can be computationally prohibitive. 

In this section, we explain how to express them as a function of simple sums such as:
\begin{equation}
\label{flowvector}
Q(A_i)\equiv \sum_{j_i\in A_i} q_i(j_i). 
\end{equation}
This reduces the number of operations from $M^n$ down to $M$ for any
order $n$.   
In the case of the analysis of anisotropic flow, $Q(A_i)$ is the usual
flow vector for particle set
$A_i$~\cite{Danielewicz:1985hn,Poskanzer:1998yz}.\footnote{Recursive
  algorithms for subtracting self-correlations in this context are
  given by Bilandzic {\it et al.\/} in Sec.~III~A of Ref.~\cite{Bilandzic:2013kga}.}

The idea is to factorize the sum, as in Eq.~(\ref{factorization}). 
In the case when there is some overlap in the sets of particles $A_i$, however, 
one must subtract terms in the sum where the same particle appears more than once, as in Eq.~(\ref{selfcorr2}). 
In the case of pair correlations, 
the notation (\ref{flowvector})  allows to recast this subtraction in compact form: 
\begin{equation}
\label{selfcorr22}
Q(A_1,A_2)=Q(A_1)Q(A_2)-Q(A_1\cap A_2). 
\end{equation}
We have introduced the auxiliary notation
\begin{equation}
\label{intersection}
Q(A_1\cap A_2)\equiv \sum_{j\in A_1\cap A_2} q_1(j)q_2(j), 
\end{equation}
where the sum runs over all particles belonging to both $A_1$ and $A_2$. 

Consider now a 3-particle correlation. The product $Q(A_1)Q(A_2)Q(A_3)$ contains all possible triplets, plus the self-correlations which must be removed. 
One separates the sum into different contributions, depending on which particles are identical:  
(1) $j_1$, $j_2$, $j_3$ all different, (2) $j_1=j_2\not= j_3$, (3) $j_2=j_3\not= j_1$, (4)  $j_1=j_3\not= j_2$,  (5) $j_1=j_2= j_3$. 
This decomposition can be represented with Young diagrams:
\begin{equation}
\label{ytableau}
\young(1,2,3)+
\young(12,3)+
\young(23,1)+
\young(13,2)+
\young(123),
\end{equation}
where each box is associated with a set $A_i$. 
In each diagram, different rows correspond to different values of the indices, and the values of the indices are identical for boxes belonging to the same horizontal row. 
The first term in Eq.~(\ref{ytableau}) is the term we want to isolate and others must be subtracted. 
The subtraction, which generalizes Eq.~(\ref{selfcorr2}) to third order, is derived in 
Appendix~\ref{s:recursion}. One obtains~\cite{Bilandzic:2013kga}:
\begin{eqnarray}
\label{n=3r}
Q(A_1,A_2,A_3)&=&Q(A_1)Q(A_2)Q(A_3)\cr&& -Q(A_1\cap A_2)Q(A_3)\cr
&&-Q(A_2\cap A_3)Q(A_1)\cr
&&-Q(A_1\cap A_3)Q(A_2)\cr
&&+2 Q(A_1\cap A_2\cap A_3), 
\end{eqnarray}
which we represent diagrammatically as 
\begin{equation}
\label{y3}
Q(A_1,A_2,A_3)=\yng(1,1,1)-
\yng(2,1)+
2 \,\yng(3),
\end{equation}
where we have omitted the labels since the weights are identical for
all permutations of the indices. 
The right-hand side of Eq.~(\ref{n=3r}) is a sum over all partitions of the set 
$\{A_1,A_2,A_3\}$. This can be generalized to arbitrary $n$, as shown in Appendix~\ref{s:recurrence}. 
The contribution of a partition is the product of contributions
of its subsets, called blocks.  
Each row in the Young diagram corresponds to a block of the partition. 
The contribution of a block of $k$ elements $\{A_{i_1},\cdots,A_{i_k}\}$ is 
the product of the integer weight  $(-1)^{k-1} (k-1)!$ and 
$Q(A_{i_1}\cap\cdots\cap A_{i_k})$, which is defined 
by a straightforward generalization of Eq.~(\ref{intersection}):
\begin{equation}
Q(A_{i_1}\cap\cdots\cap A_{i_k})\equiv\sum_{j\in A_{i_1}\cap\cdots\cap
    A_{i_k}}q_{i_1}(j)\cdots q_{i_k}(j).
\end{equation}
Blocks of  $k=1$, $2$, $3$ elements get respective weights of 
$1$, $-1$, $2$, which explains the factors in front of each term in Eq.~(\ref{n=3r}).  

We write explicitly, for sake of illustration, the corresponding formula for  the 4-particle correlation~\cite{Bilandzic:2013kga}:
\begin{eqnarray}
\label{n=4r}
Q(A_1,A_2,A_3,A_4)&=&Q(A_1)Q(A_2)Q(A_3)Q(A_4)\cr&& 
-Q(A_1\cap A_2)Q(A_3)Q(A_4)\cr&&-Q(A_1\cap A_3)Q(A_2)Q(A_4)\cr&&-
Q(A_2\cap A_3)Q(A_1)Q(A_4)\cr&&
-Q(A_1\cap A_4)Q(A_2)Q(A_3)\cr&&-Q(A_2\cap A_4)Q(A_1)Q(A_3)\cr&&-
Q(A_3\cap A_4)Q(A_1)Q(A_2)\cr&&
+Q(A_1\cap A_2)Q(A_3\cap A_4)\cr&&
+Q(A_1\cap A_3)Q(A_2\cap A_4)\cr&&
+Q(A_1\cap A_4)Q(A_2\cap A_3)\cr&&
+2 Q(A_1\cap A_2\cap A_3)Q(A_4)\cr&& 
+2 Q(A_2\cap A_3\cap A_4)Q(A_1)\cr&& 
+2 Q(A_1\cap A_3\cap A_4)Q(A_2)\cr&& 
+2 Q(A_1\cap A_2\cap A_4)Q(A_3)\cr&& 
-6 Q(A_1\cap A_2\cap A_3\cap A_4),
\end{eqnarray}
which we represent diagrammatically as:
\begin{equation}
\label{y4}
\yng(1,1,1,1)-
\yng(2,1,1)+
\yng(2,2)+
2\,\yng(3,1)-
6\,\yng(4).
\end{equation}

The weight of a given partition can be read directly from the Young diagram.  
It is the product over all rows (all blocks) of  $(-1)^{k-1} (k-1)!$, where 
$k$ is the number of boxes in the row (number of elements in block).

In order to implement the subtraction in the most general case, one must 
generate all partitions of the set $\{A_1,\cdots,A_n\}$. An efficient algorithm has been 
described by Orlov, with a public domain C++ implementation available \cite{Orlov}. A sum can then be taken, with coefficient for each term given by the above formula.

\section{From moments to cumulants}
\label{s:cumulants}

We now define the cumulants of $n$-particle correlations. 
Let $f(p_1,\cdots,p_n)dp_1\cdots dp_n$ denote the probability of finding a $n$-tuple in $dp_1\cdots dp_n$. 
We call  $f(p_1,\cdots,p_n)$ the $n$-point function. It is normalized to the average number of $n$-tuples:
\begin{equation}
\label{ntuples}
\int_p f(p_1,\cdots,p_n) = \langle M(M-1)\cdots (M-n+1)\rangle,
\end{equation}
where $\int_p$ is a shorthand notation for $\int dp_1\cdots dp_n$, and $M$ denotes the total multiplicity of an event. 

The connected $n$-point function $f_c(p_1,\cdots,p_n)$, or cumulant, 
is the contribution of the $n$-particle correlation. 
For any order $n$, it is defined recursively by isolating $n$-particle
clusters through an order-by-order decomposition of
$f(p_1,\cdots,p_n)$~\cite{kubo}. 
The 1-point functions are equal by definition: 
\begin{equation}
\label{connectedc1}
f(p_1)=f_c(p_1).
\end{equation}
The two-point function is the sum of an uncorrelated part and a
correlated part $f_c(p_1,p_2)$:
\begin{equation}
\label{connectedc2}
f(p_1,p_2)=f_c(p_1,p_2)+f_c(p_1)f_c(p_2).
\end{equation}
Similarly, one defines $f_c(p_1,p_2,p_3)$ as the part of
$f(p_1,p_2,p_3)$ which is not due to lower-order correlations~\cite{Ozonder:2016xqn}:
\begin{eqnarray}
\label{connectedc3}
f(p_1,p_2,p_3)&=&f_c(p_1,p_2,p_3)\cr
&&+f_c(p_1,p_2)f_c(p_3)\cr
&&+f_c(p_2,p_3)f_c(p_1)\cr
&&+f_c(p_1,p_3)f_c(p_2)\cr
&&+f_c(p_1)f_c(p_2)f_c(p_3).
\end{eqnarray}
The right-hand side of this equation is again a sum over partitions
of the set $\{p_1,p_2,p_3\}$, where each cluster corresponds to a block of the partition. It can be represented diagrammatically as: 
\begin{equation}
\label{ytableaubis}
\young(123)+
\young(12,3)+
\young(23,1)+
\young(13,2)+
\young(1,2,3). 
\end{equation}
Generalization to arbitrary order $n$ is straightforward. 

The average of $Q(A_1,\cdots,A_n)$ over many collision events is a weighted integral of $f(p_1,\cdots,p_n)$:
\begin{equation}
\label{Qvsfn}
\langle Q(A_1,\cdots,A_n)\rangle=\int_p q_1(p_1)\cdots q_n(p_n) f(p_1,\cdots,p_n), 
\end{equation}
We refer to such averages as {\it moments\/}. 
The cumulant decomposition applies to moments after multiplying
equations (\ref{connectedc2}), (\ref{connectedc3}) by $q_i(p_i)$ and integrating over $p_i$. 
The cumulant of order 2 is thus given by the inversion formula
\begin{equation}
\langle Q(A_1,A_2)\rangle_c\equiv\langle Q(A_1,A_2)\rangle
-\langle Q(A_1)\rangle\langle Q(A_2)\rangle,
\end{equation}
which generalizes Eq.~(\ref{cumulant2}), and which we rewrite synthetically as 
\begin{equation}
\label{newcumul2}
\langle Q_1Q_2\rangle_c=\langle Q_1Q_2\rangle
-\langle Q_1\rangle\langle Q_2\rangle.
\end{equation}
Note that the cumulant is unchanged if one shifts $Q_i$ by a constant value. This property of 
translational invariance~\cite{Teaney:2010vd}, which is true to all orders, explains why cumulants are remarkably stable 
with respect to detector imperfections. 

We write for sake of illustration the 
inversion formulas giving cumulants of order 3 and 4 as a function of the corresponding moments: 
\begin{eqnarray}
\label{cumul3}
\langle Q_1Q_2Q_3\rangle_c&=&\langle Q_1Q_2Q_3\rangle\cr
&&-\langle Q_1Q_2\rangle\langle Q_3\rangle\cr
&&-\langle Q_1Q_3\rangle\langle Q_2\rangle\cr
&&-\langle Q_2Q_3\rangle\langle Q_1\rangle\cr
&&+2\langle Q_1\rangle\langle Q_2\rangle\langle Q_3\rangle,
\end{eqnarray}
and 
\begin{eqnarray}
\label{cumul4}
\langle Q_1Q_2Q_3Q_4\rangle_c&=&\langle Q_1Q_2Q_3Q_4\rangle\cr
&&-\langle Q_1Q_2Q_3\rangle\langle Q_4\rangle\cr
&&-\langle Q_2Q_3Q_4\rangle\langle Q_1\rangle\cr
&&-\langle Q_1Q_3Q_4\rangle\langle Q_2\rangle\cr
&&-\langle Q_1Q_2Q_4\rangle\langle Q_3\rangle\cr
&&-\langle Q_1Q_2\rangle\langle Q_3Q_4\rangle\cr
&&-\langle Q_1Q_3\rangle\langle Q_2Q_4\rangle\cr
&&-\langle Q_1Q_4\rangle\langle Q_2Q_3\rangle\cr
&&+2\langle Q_1Q_2\rangle\langle Q_3\rangle\langle Q_4\rangle\cr
&&+2\langle Q_1Q_3\rangle\langle Q_2\rangle\langle Q_4\rangle\cr
&&+2\langle Q_2Q_3\rangle\langle Q_1\rangle\langle Q_4\rangle\cr
&&+2\langle Q_1Q_4\rangle\langle Q_2\rangle\langle Q_3\rangle\cr
&&+2\langle Q_2Q_4\rangle\langle Q_1\rangle\langle Q_3\rangle\cr
&&+2\langle Q_3Q_4\rangle\langle Q_1\rangle\langle Q_2\rangle\cr
&&-6\langle Q_1\rangle\langle Q_2\rangle\langle Q_3\rangle\langle Q_4\rangle.
\end{eqnarray}
The right-hand sides of these equations are again sums over partitions of the sets $\{A_1,A_2,\cdots,A_n\}$ for $n=2,3,4$. 
We represent them diagrammatically as 
\begin{eqnarray}
\langle Q_1Q_2\rangle_c&=&\yng(2)-\yng(1,1)\cr
\langle Q_1Q_2 Q_3\rangle_c&=&\yng(3)-\yng(2,1)+2\,\yng(1,1,1)\cr
\langle Q_1Q_2Q_3Q_4\rangle_c&=&\yng(4)-\,\yng(3,1)\cr
&&-\,\yng(2,2)+2\,\yng(2,1,1)-6\,\yng(1,1,1,1)
\end{eqnarray}
The weight of each term is given by a classic~\cite{kubo} M\"obius inversion
formula (see Appendix~\ref{s:recurrence}).
It is equal to 
$(-1)^{n-1} (n-1)!$, where $n$ is the number of blocks of the
 partition, i.e., the number of rows of the diagram. 

Note the striking formal analogy between the subtraction of self-correlations and the cumulant expansion. 
Both involve a summation over set partitions (so both can be generated by the same algorithm \cite{Orlov}). 
The only difference is that the weight associated with each Young diagram 
involves the number of boxes in each row for self-correlations, and the number of rows for  
the cumulant expansion.  

Finally, we point out that the corrections for self-correlations cancel to some extent in the cumulant. 
Take as an example the order 3 cumulant, given by Eq.~(\ref{cumul3}). Moments of order 2 and 3 in the right-hand side must be 
corrected for self-correlations using Eqs.~(\ref{selfcorr22}) and (\ref{n=3r}), and averaged over events. 
After summing all terms, the correction to the cumulant from self-correlation can be written as
\begin{eqnarray}
&&\langle Q(A_1\cap A_2)\rangle\langle Q(A_3)\rangle - \langle Q(A_1\cap A_2) Q(A_3)\rangle\cr
&+&\langle Q(A_1\cap A_3)\rangle\langle Q(A_2)\rangle - \langle Q(A_1\cap A_3) Q(A_2)\rangle\cr
&+&\langle Q(A_2\cap A_3)\rangle\langle Q(A_1)\rangle - \langle Q(A_2\cap A_3) Q(A_1)\rangle\cr
&+&2 \langle Q(A_1\cap A_2\cap A_3)\rangle . 
\end{eqnarray}
The first line is, up to a sign, the connected correlation between $Q(A_1\cap A_2)$ and $Q(A_3)$, which is usually much smaller than their respective magnitudes. 
This is true to all orders~\cite{Borghini:2000sa}, and the contribution to cumulants from self-correlations becomes smaller and smaller as the order 
increases when collective effects are present.\footnote{In the context of the analysis of anisotropic flow, this is consistent with the expectation that higher-order cumulants are less sensitive to nonflow effects, since nonflow effects are of the same order as self-correlations~\cite{Bhalerao:2003xf}.}
It has been checked experimentally~\cite{Abelev:2008ae,Chatrchyan:2012ta} that one obtains compatible results with and without self-correlations in the limit of large cumulant order, through an analysis of Lee-Yang 
zeros~\cite{Yang:1952be,Bhalerao:2003yq}.

Self-correlations must be subtracted on an event-by-event basis, and the number of terms increases strongly with the order $n$  (it is the Bell number), 
therefore they are a limiting factor for the calculation. 
However, it is expected that they are negligible beyond a certain order (which should explicitly be checked by doing the calculation with and without self correlations).   The cumulant expansion can then be pushed to arbitrarily high order, limited only by statistical uncertainty. 

Our definition of cumulants in this section follows from our choice of 
random variables, which are the numbers of $n$-tuples in a momentum
bin $dp_1\cdots dp_n$. As discussed in Sec.~\ref{s:pairs}, 
an alternative choice is to choose as random variables the momenta
$p_1,\cdots,p_n$ 
themselves~\cite{Bilandzic:2010jr}. 
Our method also accommodates this definition, at the expense of
a slight complication: One must normalize each moment by the
average number of $n$-tuples before taking the cumulant (as in Eqs.~(7) and (8)
of Ref.~\cite{Bilandzic:2010jr}). 
This number is obtained by repeating the calculations of
Sec.~\ref{s:selfcorrelations} with $q_k(j_k)=1$ in
Eq.~(\ref{sumalldifferent}). 
However, this complication does not appear to offer any advantage.
Our formulation has the advantage that it provides a unified framework for all analyses of correlations and fluctuations, 
as we now explain. 

\section{Applications}
\label{s:applications}

The interest of cumulants is that they subtract the effect of lower correlations, and isolate $n$-particle correlations.  
If a nucleus-nucleus collision is a superposition of a fixed number $N$ of independent nucleon-nucleon collisions, 
a moment of order $n$ scales like $N^n$, while the corresponding cumulant is proportional to $N$. 
If, on the other hand, there are collective effects in the system, 
cumulants are typically of the same order as moments, 
so that large cumulants are a clear signature of collective effects. 

By collective effect, one typically means an all-particle correlation. 
Collective effects often arise from global fluctuations, which affect the whole system. 
The fluctuation  of the total multiplicity, already mentioned in Sec.~\ref{s:pairs}, is a mundane collective effect which 
can be eliminated by a fine centrality binning~\cite{Gardim:2016nrr}. 
On the other hand, 
fixing the total multiplicity $M$ in each event also generates cumulants to all orders, but they are proportional to $M$ at any order (see Appendix~\ref{s:generating}).  As a result, working at a fixed multiplicity doesn't generate fake collective effects. 

We now describe how usual cumulant analyses can be implemented within our framework. 

\subsection{Net charge fluctuations and related studies}

Our framework, where the random variables are numbers of particles in
momentum bins (as opposed to momenta of given
particles~\cite{Bilandzic:2010jr}), naturally encompasses correlation
studies involving these numbers themselves, such as 
net charge and net baryon fluctuations. 

We first illustrate the formulas derived in previous sections by
discussing the simplest case, where 
one takes all particles from the same set, $A_1=\cdots=A_n$, and all the functions $q_i$ in Eq.~(\ref{sumalldifferent}) are equal to 1. 
In this case, the summation in this equation just counts the number of $n$-tuples of $M$ particles (cf. Eq.~(\ref{ntuples})):
\begin{equation}
\label{general}
Q(A_1,\cdots,A_n)=M(M-1)\cdots (M-n+1).
\end{equation} 
This result can be used to check the validity of
Eqs.~(\ref{selfcorr22}), (\ref{n=3r}) and (\ref{n=4r}). 
Each factor $Q(...)$ in these equations is equal to $M$, 
therefore, they reduce to:
\begin{eqnarray}
Q(A_1,A_2)&=&M^2-M\cr 
Q(A_1,A_2,A_3)&=&M^3-3 M^2+2 M\cr
Q(A_1,A_2,A_3,A_4)&=&M^4-6M^3+11M^2-6M,
\end{eqnarray}
which agree with the general result (\ref{general}) after expanding in
powers of $M$. 

Averaging $Q(A_1,\cdots,A_n)$ over events, one obtains the (unnormalized) factorial moment $F_n$, which counts the average number of $n$-tuples:
\begin{equation}
F_n\equiv \langle M(M-1)\cdots (M-n+1)\rangle
\end{equation}
If $M$ is distributed according to a Poisson distribution, $F_n=\langle M\rangle^n$. 

The cumulants as defined in Sec.~\ref{s:cumulants} are 
the {\it factorial cumulants\/} $K_n$~\cite{Dremin:1993sd} of the distribution of $M$; unlike traditional multiplicity cumulants, they vanish for $n\ge 2$  for a Poisson distribution, and are therefore automatically corrected for trivial statistical fluctuations~\cite{Bzdak:2016jxo}.\footnote{Notations are not yet standardized. Bzdak {\it et al.\/} use the notation $K_n$ for traditional cumulants, and $C_n$ for factorial cumulants~\cite{Bzdak:2016jxo}, while the STAR Collaboration~\cite{Adamczyk:2013dal} uses $C_n$ for traditional cumulants.}
In the same way, one can study correlations between multiplicities in two different rapidity windows, such as forward-backward correlations~\cite{Back:2006id,Abelev:2009ag} which have been proposed as a probe of longitudinal fluctuations~\cite{Bzdak:2012tp}. Analyses could easily be extended to higher-order cumulants~\cite{Bialas:2011xk}. 

Consider next the case where $q_i$ in Eq.~(\ref{sumalldifferent}) is
the baryon number or the electric charge.  Assuming that $q_i=\pm 1$
for all particles, the moment of order 2, as defined by Eq.~(\ref{selfcorr22}), is $Q(A_1,A_2)=\Delta N_{\rm ch}^2-M$ where $=\Delta N_{\rm ch}=\sum_i q_i$ is the net charge and $M=\sum_i (q_i)^2$ the charged multiplicity. 
Thus trivial fluctuations are again subtracted, and such observables
give more direct access to interesting quantities than the raw
distribution of $\Delta N_{\rm ch}$~\cite{Adare:2015aqk} or 
traditional cumulants~\cite{Braun-Munzinger:2016yjz}.

\subsection{Anisotropic flow}
\label{s:vnanalysis}

The analysis of anisotropic flow is one the most important practical 
applications of cumulants in high-energy physics. We recall the
definition of the relevant observables, and how they are obtained in
our framework.  

The flow paradigm~\cite{Alver:2010gr,Luzum:2011mm} states that the bulk of particle production is well described by independent-particle emission from an underlying probability distribution.   
The classic picture is that of a relativistic fluid --- near
freeze-out, the system is a fluid, composed of well-defined particles
that are uncorrelated with each other (e.g., as in a Boltzmann
description).   In
other words, if every event had the same hydrodynamic initial
conditions (and a fixed orientation), the connected correlations
defined by Eqs.~\eqref{connectedc2}, \eqref{connectedc3} would
vanish.

In a single event, then, all information is contained in the single-particle distribution $F(p)$.  Note that this single particle distribution is different from $f(p)$ in Eq.~\eqref{connectedc1}, which is $F(p)$ averaged over events (and azimuthal orientation).
Anisotropic flow is the particular observation that the single-particle distribution depends on azimuthal angle $\phi$:
\begin{equation}
\label{defvn}
F(p) = \frac {N} {2\pi} \sum_{n=-\infty}^{\infty} V_n e^{in\phi},
\end{equation}
with $V_0=1$, $V_{-n}=V_n^*$, and $N$ is the average number of particles in the event. 
One denotes by $v_n\equiv |V_n|$ the anisotropic flow coefficient in harmonic $n$~\cite{Voloshin:1994mz} in a particular event.  

One can instead absorb the factor of $N$ into the coefficients~\cite{Mazeliauskas:2015vea}
\begin{equation}
F(p) = \frac {1} {2\pi} \sum_{n=-\infty}^{\infty} {\cal V}_n e^{in\phi},
\end{equation}
where now ${\cal V}_n$ = $N V_n$.  While less standard, these are convenient quantities since they are additive with respect to rebinning in momentum space, and are more natural quantities with respect to the cumulant analysis.

The single-event distribution $F(p)$ fluctuates from one event to the next.  Upon averaging over events (and fluctuations), this generates correlations to all orders.
Moments and cumulants of multiparticle correlations are 
moments and cumulants of $F(p)$.
\footnote{
There are non-trivial mathematical consequences to having uncorrelated particles in each event, e.g., for pair correlations~\cite{Gardim:2012im,Bhalerao:2014mua}.}

Even if the bulk of particle production can be described according to
this single-particle distribution, it is expected for there to be
small correlations that cannot be captured by $F(p)$.    Such effects
are typically referred to as ``non-flow'', and can arise from sources
such as Bose-Einstein correlations, resonance decays, unthermalized jets, momentum conservation, etc.
Observables involving higher-order cumulants are expected to be less
sensitive to such nonflow effects~\cite{Borghini:2000sa}. 

The measure of $v_n$ from the cumulant of order $2k$ is usually noted $v_n\{2k\}$, and it can easily be obtained with the procedure outlined above.   This generalizes the discussion of pair correlations  outlined in Sec.~\ref{s:pairs}. 
One now takes $2k$-tuples, again taken from the same region of phase space (``integrated'' flow). 
The factor $q_i$ in Eq.~(\ref{sumalldifferent}) is equal to $e^{in\phi}$ for the first $k$ particles and to $e^{-in\phi}$ for the next $k$ particles. 
The subtraction of self correlations carried out in Sec.~\ref{s:selfcorrelations} generalizes that of Ref.~\cite{Bilandzic:2010jr}. 
For instance, one easily checks that the order 4 result Eq.~(\ref{n=4r}) reproduces the numerator of Eq.~(18) of Ref.~\cite{Bilandzic:2010jr}.\footnote{
The weight of each event is the same as in Ref.~\cite{Bilandzic:2010jr}, in the sense that all $n$-tuples are treated on an equal footing (and as a result, each \textit{event} can have significantly different contribution, if multiplicity fluctuates). 
A variety of prescriptions are found in the literature. The 
scalar-product analysis~\cite{Adler:2002pu} and the Lee-Yang 
zero analysis~\cite{Bhalerao:2003xf} use a prescription equivalent to
ours. 
The cumulant analysis of Ref.~\cite{Borghini:2001vi} weights each
event by $1/M$, and 
the analysis of event-plane correlations~\cite{Aad:2014fla} implements
a similar weight $1/E_T$, where $E_T$ is the energy deposited in the
calorimeter. 
Finally, factors of $1/\sqrt{M}$ were used in the cumulant analysis of 
Ref.~\cite{Borghini:2000sa} and in studies of the distribution of the
flow vector~\cite{Adler:2002pu}).}
We denote by $\langle |Q_n|^{2k}\rangle_c$, the cumulant obtained
after subtracting self-correlations, averaging over events and
subtracting lower-order moments (Sec.~\ref{s:cumulants}). 
As explained above, it is a cumulant of the distribution of 
${\cal V}_n$. We thus define: 
\begin{equation}
{\cal V}_n\{2k\}^{2k}\equiv\frac{1}{a_{2k}} \langle |Q_n|^{2k}\rangle_c,
\end{equation}
where $a_{2k}$ is the ratio of the coefficients of the expansions of $\ln I_0(x)$ (cumulants) and $I_0(x)$ (moments) to order $x^{2k}$~\cite{Borghini:2000sa}:
\begin{equation}
\ln I_0(x)=\sum_{k=1}^{\infty} \frac{a_{2k} x^{2 k}}{2^{2k} (k!)^2},
\end{equation}
which gives $a_2=1$, $a_4=-1$, $a_6=4$, $a_8=-33$, $a_{10}=456$, $a_{12}=-9460$, etc.
This normalization ensures that ${\cal V}_n\{2k\}=|{\cal V}_n|$ for all $k$ if $|{\cal V}_n|$ were the same for all events. 

In order to match the usual normalization convention of the flow
coefficients $v_n$, one normalizes this cumulant by the average number of 
$2k$-tuples. 
The standard azimuthal cumulants are thus given by
\begin{equation}
\label{integrated}
v_n\{2k\}^{2k}\equiv\frac{1}{a_{2k}}\frac{\langle |Q_n|^{2k}\rangle_c}{F_{2k}}.
\end{equation}
This normalization ensures that again $v_n\{2k\}=v_n$ when there are no fluctuations.

For the traditional measurement of ``differential'' flow 
\cite{Borghini:2000sa}, the only difference with the previous case is that the first particle is taken from a different (restricted) phase space window $B$, while the $2k-1$ remaining particles are taken from the same set of particles $A$.  
One then scales the resulting cumulant with the corresponding number of $2k$-tuples (with 1 particle in $B$ and $2k-1$ particles in $A$).
The flow in $B$, traditionally denoted by $v'_n\{2k\}$, is again given by an equation similar to (\ref{integrated}), where one replaces in the left-hand side  $v_n\{2k\}^{2k}$ with $v'_n\{2k\}v_n\{2k\}^{2k-1}$. Note that the coefficients $a_k$ are the same as for integrated flow.

In principle, each particle can be taken from an arbitrary region in momentum space, and many other differential analyses are 
possible~\cite{Jia:2017hbm,inpreparation}.  With our normalization conventions, all differential cumulants are additively related to integrated measurements.

\subsection{Symmetric cumulants, event-plane correlations}

The normalized symmetric cumulant ${\rm NSC}(m,n)$ with $n\neq m$ can
be defined by~\cite{ALICE:2016kpq}\footnote{As with previous
  cumulants, the ALICE Collaboration chose to normalize each
  individual moment by $F_2$ or $F_4$.  See the discussion at the end
  of Sec.~\ref{s:cumulants}. We choose to have no normalization factors in the normalized measurement.}  
\begin{align}
\label{sc}
{\rm NSC}(m,n)&=\frac {\left\langle Q_n Q_m Q_{-n} Q_{-m} \right\rangle_c} {\left\langle |Q_n|^2 \right\rangle_c \left\langle |Q_m|^2 \right\rangle_c} \\
\label{scflow}
&\stackrel{\rm{(flow)}}{=}\frac{\langle |{\cal V}_m|^2 |{\cal V}_n|^2\rangle-\langle |{\cal V}_m|^2\rangle\langle |{\cal V}_n|^2\rangle}{\langle |{\cal V}_m|^2\rangle\langle |{\cal V}_n|^2\rangle}.
\end{align}
The numerator is a $4$-particle cumulant~\cite{Bilandzic:2013kga} with $q_1=e^{im\phi}$,
$q_2=e^{in\phi}$, $q_3=e^{-im\phi}$, $q_4=e^{-in\phi}$.
The fact that it is measured through a 4-particle cumulant guarantees that nonflow effects are small. 
Equation~\eqref{scflow} is the value assuming the flow paradigm of
independent particles in each event. 

The mean square values $\langle |{\cal V}_n|^2\rangle=F_2 v_n\{2\}^2$ in the denominator are 2-particle cumulants  obtained as described in 
Sec.~\ref{s:vnanalysis}, 
which may be biased by short-range correlations (especially nonflow effects) unless a rapidity gap is applied. 
The existing analysis~\cite{ALICE:2016kpq} implements a gap in the denominator, but not in the numerator, hence neglecting the effect of longitudinal fluctuations. 
it would be interesting to redo the analysis by implementing the same gap in the numerator and the denominator, and studing how ${\rm NSC}(m,n)$ varies with the gap. 

Event-plane correlations~\cite{Aad:2014fla} are Pearson correlation coefficients between  different complex flow hamonics~\cite{Luzum:2012da,Yan:2015jma}.  For instance, the two-plane correlation between ${\cal V}_2$ and ${\cal V}_4$ is defined as (we use the notation of ATLAS):
\begin{align}
\label{42correlation}
\langle\cos(4(\Phi_2-\Phi_4))\rangle_w&\equiv \frac{\left\langle Q_4 Q_{-2} Q_{-2} \right\rangle_c} {\sqrt{\left\langle |Q_4|^2\right\rangle_c\left\langle |Q_2|^4\right\rangle }} \\
&\stackrel{\rm{(flow)}}{=}\frac{\langle {\cal V}_4({\cal V}_2^*)^2\rangle}{\sqrt{\langle |{\cal V}_4|^2\rangle\langle |{\cal V}_2|^4\rangle}}.
\end{align}
The numerator is a 3-particle cumulant obtained with $q_1=e^{4i\phi}$, $q_2=q_3=e^{-2i\phi}$.
The ATLAS analysis uses two sets of particles separated with a rapidity gap, where particles 2 and 3 belong to the same bin. 
Thus self-correlations between particles 2 and 3 are not removed. It has been argued that 
they are small~\cite{Bhalerao:2013ina}, but it will be interesting to check experimentally. 
The denominator of Eq.~(\ref{42correlation}) involves moments which, as in the case of symmetric cumulants, may be biased by nonflow effect unless a rapidity gap is applied (note that the 4-particle $v_2$ factor is a moment, not a cumulant). 

It has been pointed 
out~\cite{Giacalone:2016afq} that the value of ${\rm NSC}(4,2)$ measured by ALICE seems large compared to what one would expect based on the corresponding event-plane correlation measured by ATLAS~\cite{Aad:2014fla}, where a large rapidity gap is implemented. It would be interesting to measure both the symmetric cumulant and the event-plane correlation with the same kinematic cuts. 

\subsection{Correlation between transverse momentum and anisotropic flow}

The correlation between transverse momentum and anisotropic flow recently proposed by Bozek as a further test of hydrodynamic behavior~\cite{Bozek:2016yoj} is a straightforward application of our formalism. It is a 3-particle cumulant with $q_1=p_t$, $q_2=e^{in\phi}$, $q_3=e^{-in\phi}$. Bozek recommends to use particles from three different rapidity intervals $A,B,C$ separated with gaps in order to avoid nonflow correlations but predicts that results should be identical if $A=B=C$ provided that self-correlations are subtracted.

\section{Conclusion}

We have proposed a new, unified framework for cumulant analyses which is more systematic and flexible than existing frameworks, and discussed its practical implementation for the analysis of factorial cumulants and anisotropic flow. 
A major improvement is that one can correlate particles in arbitrary regions of phase space. 
Application to proton-nucleus and nucleus-nucleus at RHIC and the LHC should shed light on longitudinal fluctuations. 
Our procedure is systematic and can be carried out to arbitrarily
large orders, which is important in order to probe collective
behavior. In particular, we have argued that the subtraction of
self-correlations, which is the limiting factor when going to higher
orders, becomes a negligible correction at large orders. This can be 
checked explicitly, and the cumulant expansion can then be extended to higher
orders. 
Finally, our new framework also extends beyond the analysis of anisotropic flow, and we anticipate a rich program of 
generalized cumulant analyses on this basis in the near future.

\section*{Acknowledgments}
This work is funded under the USP-COFECUB project Uc Ph 160-16
(2015/13) and under the FAPESP-CNRS project 2015/50438-8.  
PDF is supported by the NSF Grant DMS-1301636 and the Morris and Gertrude Fine endowment. 
MG is supported under DOE Contract No. DE-SC0012185 and the Welch Foundation (Grant No. C-1845), and 
thanks Wei Li  and Hubert Hansen for discussions. 
JYO thanks the Department of Theoretical Physics, Mumbai, where this paper was written, for hospitality and financial support. 
We thank Giuliano Giacalone for careful reading of the manuscript. 

\appendix
\section{M\"obius inversion}
\label{s:recurrence}
In this Appendix, we recall some known facts~\cite{stanley} on M\"obius inversion applied to functions
of set partitions. We show in particular how it implies the moment/cumulant relations of Sec.~\ref{s:cumulants},
as well as the self-correlation subtraction of Section~\ref{s:selfcorrelations}.

\subsection{Set partitions and M\"obius inversion}
Let $[n]\equiv\{1,2,...,n\}$ and ${\mathcal P}(n)$ the set of its
partitions. We denote by ${\mathbf  I}=\{I_1,...,I_k\}$ a partition, where 
$I_j$ are subsets of $[n]$, called blocks of the partition. 
The number of blocks $k$ is called the length of the partition ${\mathbf I}$ and
denoted by $|{\mathbf I}|$.  
There is a natural partial order relation on partitions which we
denote by $\leq$ and is defined as follows. 
If ${\mathbf I}$ and ${\mathbf J}$ are two partitions of $[n]$, 
${\mathbf I}\leq {\mathbf J}$
if the partition ${\mathbf I}$ is a refinement of the partition
${\mathbf J}$, that is, 
if each block of ${\mathbf I}$ is included in a block of ${\mathbf J}$. 
For instance we have $\{\{1,2\},\{3\},\{4\}\}\leq \{ \{1,2\},\{3,4\}\}$ but the order relation does not relate $\{\{1,2\},\{3,4\}\}$
to $\{\{1,3\},\{2,4\}\}$. 
The finest partition is denoted by 
$\hat{\mathbf 0}\equiv\{\{1\},\{2\},...,\{n\}\}$ and the coarsest partition by
$\hat{\mathbf 1}\equiv{[n]}$.
For all ${\mathbf I}\in {\mathcal P}(n)$, 
$\hat{\mathbf 0}\leq {\mathbf I}\leq \hat{\mathbf 1}$.

The M\"obius inversion formula goes as follows~\cite{peccati}. 
Assume we have two real functions $f,g$ defined on  ${\mathcal P}(n)$, such that for all ${\mathbf I}\in {\mathcal P}(n)$:
\begin{equation}\label{moeb}
 f({\mathbf J})=\sum_{\hat{\mathbf 0}\leq {\mathbf I}\leq {\mathbf J}}
 g({\mathbf I}).
\end{equation}
then we have the inverse relation:
\begin{equation}\label{invmoeb}
g({\mathbf J})=\sum_{\hat{\mathbf 0}\leq {\mathbf I}\leq {\mathbf J}}
\mu({\mathbf I},{\mathbf J})\, f({\mathbf I}),
\end{equation}
where the M\"obius function $\mu({\mathbf I},{\mathbf J})$ is
given by 
\begin{equation}
\mu({\mathbf I},{\mathbf J})=(-1)^{|{\mathbf I}|-|{\mathbf J}|}
\prod_{i=1}^n ((i-1)!)^{r_i({\mathbf I},{\mathbf J})},
\end{equation} 
where $r_i({\mathbf I},{\mathbf J})$ denotes the number of blocks of
$\mathbf J$ containing exactly $i$ blocks of $\mathbf I$. 

Likewise, there is a {\it dual} M\"obius inversion formula: assuming $f,g$ are related via
\begin{equation}\label{dmoeb} f({\mathbf I})=\sum_{{\mathbf I}\leq {\mathbf J}\leq \hat{\mathbf 1}} g({\mathbf J})\end{equation}
for all ${\mathbf I}\in {\mathcal P}(n)$
then we have the inverse relations:
\begin{equation}\label{dinvmoeb}g({\mathbf I})=\sum_{{\mathbf I}\leq {\mathbf J}\leq \hat{\mathbf 1}} \mu({\mathbf I},{\mathbf J})\, f({\mathbf J})\end{equation}

\subsection{Applications}

An example of the decomposition (\ref{moeb}) is the decomposition of
moments into cumulants (Section \ref{s:cumulants}). 
One defines $f({\mathbf I})$ as the product of moments over all
blocks, and $g({\mathbf I})$ as the product of cumulants over all
blocks, that is; 
\begin{eqnarray}
f({\mathbf I})&\equiv&\prod_{j=1}^{|{\mathbf I}|}\left\langle \prod_{l\in I_j} Q_l\right\rangle\cr
g({\mathbf I})&\equiv&\prod_{j=1}^{|{\mathbf I}|}\left\langle \prod_{l\in I_j} Q_l\right\rangle_c.
\end{eqnarray}
Then the M\"obius inversion formula \eqref{invmoeb} specialized to
${\mathbf J}=\hat{\mathbf 1}$ 
gives the connected correlation function 
$\langle Q_1\cdots Q_n\rangle_c$ in terms of moments
as a sum over partitions, with coefficients
\begin{equation}
\mu({\mathbf I},\hat{\mathbf 1})=(-1)^{|{\mathbf I}|-1}\, (|{\mathbf I}|-1)!
\end{equation}
Formulas (\ref{newcumul2}), (\ref{cumul3}), (\ref{cumul4}) follow as well as the general case.

Similarly, the dual M\"obius inversion formula applies directly to the problem of subtracting self-correlations (Section \ref{s:selfcorrelations}).
We now define 
$f([n])\equiv Q(A_1)...Q(A_n)$ (sum over all indices, including self
correlations) 
$g([n])\equiv Q(A_1,\cdots,A_n)$ (sum over different indices), and
more generally: 
\begin{eqnarray}
f({\mathbf I})&\equiv& Q(\cap_{i\in I_1}A_i)Q(\cap_{i\in I_2}A_i)...
Q(\cap_{i\in I_n}A_i)\cr
g({\mathbf I})&\equiv& Q(\cap_{i\in I_1}A_i, \cap_{i\in
  I_2}A_i,...,\cap_{i\in I_n}A_i).
\end{eqnarray}
The relations \eqref{dmoeb} are clearly satisfied,
as shown by splitting uncontrained sums into sums of constrained ones,
for instance:
\begin{eqnarray}
Q(A_1)Q(A_2)&=&\sum_{i,j} q_1(i)q_2(j)\cr
&=&\sum_{i\neq j}q_1(i)q_2(j)+\sum_{i=j} q_1(i)q_2(i)\cr
&=&Q(A_1, A_2)+Q(A_1\cap A_2).
\end{eqnarray} 
The dual M\"obius inversion formula
\eqref{dinvmoeb} for ${\mathbf I}=\hat{\mathbf 0}$ gives the expression for $g(\hat{\mathbf 0})=Q(A_1,A_2,...,A_n)$ in terms of the $f({\mathbf J})$,
with coefficients 
\begin{equation}
\mu(\hat{\mathbf 0},{\mathbf J})=\prod_{i=1}^{|{\mathbf J}|}
(-1)^{|J_i|-1}\, (|J_i|-1)! 
\end{equation}
where $|J_i|$ is the cardinality of the block $J_i$, equal to the length of the $i$-th row of the Young diagram of ${\mathbf J}$.
Formulas (\ref{selfcorr22}), (\ref{n=3r}), (\ref{n=4r}) and their generalization follow.

\section{Recursion relation}
\label{s:recursion}

We derive a relation for generating the self-correlation corrections
order by order. 
It is simple to implement and easy to understand, but less efficient numerically than the general method exposed in Sec.~\ref{s:selfcorrelations}.
Similar relations have been previously derived in
Ref.~\cite{Bilandzic:2013kga}.

We want to evaluate sums of the type 
\begin{equation}
\label{defsum}
Q(A_1,\ldots,A_n)\equiv\sum_{j_1\in A_1,\ldots,j_n\in
 A_n}q_1(j_1)\ldots q_n(j_n),
\end{equation} 
where all indices in the sum are different. 
Self correlations can be subtracted order by order. 
Once one has a formula that works for $n-1$, then to order $n$ one has
\begin{eqnarray}
\label{recurrence}
Q(A_1,\ldots,A_n)&=&Q(A_1,\ldots,A_{n-1})Q(  A_n)\cr
&&-
\sum_{k=1}^{n-1}Q(A_1,...,A_k\cap A_n,...,A_{n-1}).\cr
\end{eqnarray} 
The first term in the right-hand side takes into account the
conditions that the indices $i_1$ to $i_{n-1}$ are all different, and
the last term subtracts the contributions from $i_k=i_n$. 
It is understood that the product $q_iq_j$ should be used
together  with the intersection $A_i\cap A_j$. 
For $n=2$, Eq.~(\ref{recurrence}) gives Eq.~(\ref{selfcorr2}).
For $n=3$, it gives
\begin{eqnarray}
\label{n=3}
Q(A_1,A_2,A_3)&=&Q(A_1,A_2)Q(A_3)-Q(A_1\cap A_3,A_2)\cr&&-Q(A_1,A_2\cap A_3).
\end{eqnarray}
Substituting Eq.~(\ref{selfcorr2}) into Eq.~(\ref{n=3}), one obtains Eq.~(\ref{n=3r}).
For $n=4$, Eq.~(\ref{recurrence}) gives
\begin{eqnarray}
\label{n=4}
Q(A_1,A_2,A_3, A_4)&=&Q(A_1,A_2,A_3)Q(A_4)\cr&&-Q(A_1\cap A_4,A_2,A_3)\cr
&&-Q(A_1,A_2\cap A_4,A_3)\cr
&&-Q(A_1,A_2,A_3\cap A_4)
\end{eqnarray}
Substituting Eq.~(\ref{n=3}) into Eq.~(\ref{n=4}), one obtains  Eq.~(\ref{n=4r}).
Generating the subtraction to order $n$ with this method requires $n!$ operations so that it is less efficient in practice than generating partitions, since the number of partitions is the Bell number which grows more slowly with $n$ than $n!$. 

\section{Generating function}
\label{s:generating}

Moments and cumulants to all orders are conveniently expressed in terms of generating functions. 
The moment defined by Eq.~(\ref{Qvsfn}) can be obtained 
 by expanding the generating function:
\begin{equation}
\label{generating}
G(z_1,\cdots,z_n)\equiv\left\langle \prod_{j=1}^M\left(1+z_1 q_1(j)+\cdots+z_n q_n(j)\right)\right\rangle,
\end{equation}
where the product runs over all particles in the event. 
If one expands the product, the moment defined by (the average over events of) Eq.~(\ref{sumalldifferent}) is the coefficient in front of $z_1\cdots z_n$. 
The corresponding cumulant is given by the expansion of $\ln G$ to the same order:
\begin{equation}
\langle Q_1\cdots Q_n\rangle_c=\left.\frac{\partial^n}{\partial z_1\cdots\partial z_n}\ln G(z_1,\cdots,z_n)
 \right|_{z_1=\cdots=z_n=0}.
\end{equation}
Writing $G$ as a product over all particles~\cite{Borghini:2001vi} guarantees that self-correlations do not appear at any order in the expansion. 
If one does not subtract self correlations, then the generating function takes the usual exponential form~\cite{Borghini:2000sa}:
\begin{equation}
\label{generatingb}
G(z_1,\cdots,z_n)\equiv\left\langle \exp\left( z_1 Q(A_1)+\cdots+z_n Q(A_n)\right)\right\rangle,  
\end{equation}
with $Q(A_i)=\sum_j q_i(j)$. 
Note that Eq.~(\ref{generating}) can also be written in an exponential form analogous to (\ref{generatingb}) by introducing Grassmann variables~\cite{Creutz:1992ex}. This formal analogy shows that the algebraic relations linking moments to cumulants, derived in Sec.~\ref{s:cumulants}, are identical irrespective of whether or not self-correlations are subtracted. 
Both forms of the generating function have been used~\cite{Abelev:2008ae,Chatrchyan:2012ta} in the context of the analysis of elliptic flow with Lee-Yang zeros~\cite{Bhalerao:2003xf}.

As an application of the formalism, we evaluate the generating function (\ref{generating}) in the simple case of independent particles, where all terms in the product are independent. 
First, consider the case where the multiplicity $M$ is fixed. Then, Eq.~(\ref{generating}) gives
\begin{equation}
\label{fixedm}
G(z_1,\cdots,z_n)=\left(1+z_1 \langle q_1\rangle+\cdots+z_n \langle q_n\rangle\right)^M.
\end{equation}
Therefore, $\ln G$ is proportional to $M$ and cumulants of arbitrary order scale like $M$, as stated in 
the first paragraph of Sec.~\ref{s:applications}. 

If, on the other hand, all the connected $n$-point functions $f(p_1,\cdots , p_n)$ vanish for $n\ge 2$, then, 
the multiplicity $M$ follows a Poisson distribution:
\begin{equation}
\label{poisson}
p_M=\frac{\langle M\rangle^M}{M!} e^{-\langle M\rangle}. 
\end{equation}
Inserting into Eq.~(\ref{generating}) and summing the series, one obtains
\begin{equation}
\label{summation}
G(z_1,\cdots,z_n)=\exp\left( \langle M\rangle (z_1 \langle q_1\rangle+\cdots+z_n \langle q_n\rangle)\right).
\end{equation}
Therefore, $\ln G(z_1,\cdots,z_n)$ is linear in all the variables $z_1$ and cumulants  of order $\ge 2$ vanish identically, as expected.

\end{document}